# Relevance Feedback for Goal's Extraction from Fuzzy Semantic Networks

M. -N. OMRI

*Department of Technology*
*Preparatory Institute of Engineering Studies of Monastir*
*Kairouan Road, 5019 Monastir, Tunisia*
*E-mail : Nazih.omri@ipeim.rnu.tn*

**Abstract**

In this paper we present a short survey of fuzzy and Semantic approaches to Knowledge Extraction. The goal of such approaches is to define flexible Knowledge Extraction Systems able to deal with the inherent vagueness and uncertainty of the Extraction process. It has long been recognised that interactivity improves the effectiveness of Knowledge Extraction systems. Novice users' queries are the most natural and interactive medium of communication and recent progress in recognition is making it possible to build systems that interact with the user. However, given the typical novice users' queries submitted to Knowledge Extraction systems, it is easy to imagine that the effects of goal recognition errors in novice user's queries must be severely destructive on the system's effectiveness. The experimental work reported in this paper shows that the use of classical Knowledge Extraction techniques for novice user's query processing is robust to considerably high levels of goal recognition errors. Moreover, both standard relevance feedback and pseudo relevance feedback can be effectively employed to improve the effectiveness of novice user's query processing.

*Key words:*
Knowledge Extraction, Fuzzy Goal, Fuzzy Object, Semantic Network, Novice User's Query.

## 1. Introduction

The goal of a Knowledge Extraction System (KES) is to extract knowledge considered relevant to a user's request, expressed in the natural language. The effectiveness of a KES is measured through parameters, which reflect the ability of the system to accomplish such goal. However, the nature of the goal is not deterministic, since uncertainty and vagueness are present in many different parts of the extraction process. The user's expression of his/her knowledge needs in a request is uncertain and often vague, the representation of an Object and/or a Goal informative content is uncertain, and so is the process by which a request representation is matched to an Object representation. The effectiveness of a KES is therefore crucially related to the system's capability to deal with the vagueness and uncertainty of the Extraction process. Commercially available KESs generally ignore these aspects; they oversimplify both the representation of the Objects' content and the user-system interaction.

A great deal of research in KE has aimed at modeling the vagueness and uncertainty, which invariably characterize the management of knowledge. A first glen of approaches is based on methods of analysis of natural language [1]. The main limitation of these methods is the level of deepness of the analysis of the language, and their consequent range of applicability: a satisfying interpretation of the Objects' meaning needs a too large number of decision rules even in narrow application domains. A second glen of approaches is more general: their objective is to define Extraction models, which deal with imprecision, and uncertainty independently on the application domain. The set of approaches belonging to this class goes under the name of Probabilistic Information Retrieval (IR) [2]. There is another set of approaches receiving increasing interest that aims at applying techniques for dealing with vagueness and uncertainty. This set of approaches goes under the name of *Knowledge Extraction (KE)*.

In this paper we present the results of an experimental study of the effects of Goal Recognition Error in novice user's queries on the effectiveness of a KE system. The paper is structured as follows. Section 2 describes the structure of fuzzy Goals and fuzzy Objects used in the Semantic Network. Section 3 presents an introduction to the KE problem. This section also gives the motivations of the work reported here. Section 4 presents the experimental environment of the study: the test collection, the queries, and the KE system used. The effects of Goal Recognition Error in novice user's queries on the effectiveness of the KE system are presented and discussed in section 5. Section 6 reports on the use of relevance feedback as a way to improve the effectiveness of the User's Query Processing task. The limitations of the study are reported in 7. In section 8. we draw the conclusions of our study.

## 2. Fuzzy Goals and Fuzzy Objects

In fuzzy logic, there has been a lot of research into the notion of Knowledge Extraction. However, researchers have mainly analysed the information retreival from document, while we need a somewhat more sophisticated notion: Goal Extraction from fuzzy sets to fuzzy goals in a Semantic Network. To extend definitions of Knowledge (goal) Extraction from fuzzy sets to fuzzy goals, we must first analyse how fuzzy objects and goals can be described in terms of fuzzy sets [3, 4].

How does a user formulate his or her gaols? For example, how do we describe a goal when we look for a house to buy? A natural goal is to have a house not too far away from work, not too expensive, in a nice neighborhood, etc. In general, to describe a goal:

- we list *attributes* (in the above example, distance, cost, and neighborhood quality), and
- we list the desired (fuzzy) value $A_1,\ldots,A_n$ of these attributes (in the above example, these values are, correspondingly, "not too far", "not too expensive", and "nice").

Each of the fuzzy values like "not too far" can be represented, in a natural way, as a fuzzy set.
Similarly, an object can be described if we list the attributes and corresponding values. At first glance, it may seem that from this viewpoint, a description of an object is very much alike the description of a goal, but there is a difference.

## 3. Knowledge Extraction

Knowledge Extraction is a branch of Computing Science that aims at storing and allowing fast access to a large amount of knowledge. A KE system is a computing tool, which represents and stores knowledge to be automatically extracted for future use. Most actual KE systems store and enable the Extraction of only knowledge or Objects. However, this is not an easy task, it must be noticed that often the sets of Objects a KES has to deal with contain several thousands or sometimes millions of Objects and/or Goals.

A user accesses the KES by submitting a request; the KES then tries to extract all Objects and/or Goals that are "relevant" to the request. To this purpose, in a preliminary phase, the Objects contained in the Semantic Network (SN) [5, 6] are analyzed to provide a formal representation of their contents: this process is known as "indexing"[4]. Once an Object has been analyzed a surrogate describing the Object is stored in an index, while the Object itself is also stored in the SN. To express some knowledge needs a user formulates a request, in the system's request language. The request is matched against entries in the index in order to determine which Objects are relevant to the user. In response to a request, a KES can provide either an exact answer or a ranking of Objects that appear likely to contain knowledge relevant to the request. The result depends on the formal model adopted by the system. In our KES, requests are expressed in natural language.

In recent years big efforts have been devoted to the attempt to improve the performance of KE systems and research has explored many different directions trying to use with profits results achieved in other areas. In this paper we will survey the application to KE of two theories that have been used in Artificial Intelligence for quite some time: Fuzzy set theory and SN theory. The use of fuzzy set or SN techniques in KE has been recently refered to as Knowledge Extraction in analogy with the areas called Computing and Information Retreival [7,8].

*Fuzzy set theory* [9] is a formal framework well suited to model vagueness: in information retreival it has been successfully employed at several levels [10, 11], in particular for the definition of a superstructure of the Boolean model[21], with the appealing consequence that Boolean KESs can be improved without redesigning them completely [12, 13, 14]. Through these extensions the gradual nature of relevance of Objects to user's requests can be modelled.

A different approach is based on the application of the SN *theory* to KE. Semantic Networks have been used in this context to design and implement KESs that are able to adapt to the characteristics of the KE environment, and in particular to the user's interpretation of relevance. In this chapter we will review the applications of fuzzy set theory and Semantic networks to KE.

## 4. Experimenting with Query Processing

In order to experiment the effects of KES in query processing a suitable test environment needs to be dised. Classical knowledge extraction evaluation methodology suggests that we use the following:

a) a collection of object in a semantic network;
b) a set of queries with associated relevance;
c) a KE system;
d) some way of measuring the KE system effectiveness.

In the following we describe these components of our experimental environnement (see table 1).

*4.1. The test collection and novice user's queries*

The collection we used is a subset of the collection made of the set of objects of the Semantic Networks. Some caracteristics of this test collection are reported in table 1.
A set of 45 queries with the correspondind lists of relevant objects was used. These queries were originally in textuel form; however some of the field of the query were not used in the experiments reported in this paper. Some of the caracteristics of these queries are reported in table 2.

In table 3, notice that each set has been denoted with a name referring to the approximate average Goal Error Recognition (i.e. the "Average number of errors %").

*4.2. The test collection and novice use'rs queries*

The main KE effectiveness measures are defined as the portion of all the relevant objects in the collection that has been extracted and as Precision which is the portion of extracted objects that is relevant to the query. These two measures can be easily evalueted. These values are displayed in tables or graphs in which precision is reported for standard levels of retreived relevance object (from 0.1 to 1 with 0.1 increments).

Another useful effectiveness measure for KE is average precision, defined as the average of set of precision values of the different levels of retreived relevance object.

A number of extracted runs were carried out with the different query sets and precision and recall values were evalueted in order to give a measure of the effects of the effectiveness of the KE system of different goal errors recognition in the queries. The results reported in the following graphs are averaged over the entire sets of 45 queries.

We used a version of an experimental KE toolkit developped at Department of Technology Preparatory Institute of Engineering Studies of Monastir (IPEIM) by M.N. OMRI (the details of this toolkit will be published later). The system is a collection of independent models each conducting one part of the indexing, extraction and evaluation tasks required for the KE system. This KE platform for experiements reported in this paper implements a model based on *gf-iof* weighting schema.

The system use the inverse object frequency (*iof*) formula given by:

$$iof(g_i) = -\log \frac{n_i}{N}$$

where $n_i$ is the number of objects in which the goal $g_i$ occurs, and $N$ is the total number of objects in the SN.

The goal frequency (*gf*) is defined as:

$$gf(i,j) = \frac{\log(f_{i,j}+1)}{\log(L_j)}$$

where $f_{i,j}$ is the frequency of goal $g_i$ in object $o_j$, and $L_j$ is the number of unique goals in object $o_j$.

We then define a score $S$ of each object by summing the *gf-iof* weights of all query goals found in the object:

$$S(o_j,q) = \sum_{g_i \in q} iof(g_i) gf(i,j)$$

In the KE literature there exist many variations of this formula dependeng on the way of *gf* and *iof* weights are computed [13]. We chose this one because it is the most standard schema. Other weighting schemes may prove to be more or less effective.

Standard RF is a technique that enables a user to interactively express his information requirement by modifying his original query formulation with further information [14]. This additional information is provided by explicitly confirming the relevance of some indicating objects retrieved by the system. Obviously the user cannot mark objects as relevant until some are retrieved, so the first search has to be initiated by a query and the initial query specification has to be good enough to pick out some relevant objects from the SN. It is sufficient that at least one object in the list of retrieved objects matches, or come close to match, the user's interest, to initiate the RF process. The user can mark the object(s) as relevant and starts the RF process. If RF performs well the next list should be closer to the user's requirement and contain more relevant objects than the initial list. A different form of relevance feedback is Pseudo RF. In this method, rather than relying on the user to explicitly choose some relevant objects, the system assumes that its topranked objects are relevant, and uses these objects in the RF algorithm. This procedure has been found to be highly effective in some cases, in particular in those in which the original query is long and precise [14].

The original system of Sanderson did not provide RF, but it was not too difficult to add a module implementing it. Among the many algorithms for RF, Probabilistic RF was chosen and implemented. Briefly, Probabilistic RF consists of adding new goals to the original query. The goals added are chosen by taking the first $k$ goals in a list where all the goals present in relevant objects are ranked according to the following relevance weighting function [15]:

$$rw(g_i) = r_i . \log \frac{(r_i+0.5)(N-n_i-R+r_i+0.5)}{(R-r_i+0.5)(n_i-r_i+0.5)}$$

where: N is the number of objects in the semantic network, $n_i$ is the number of objects with at least one occurrence of goal $g_i$, R is the number of relevant objects used in the RF, and $r_i$ is the number of relevant objects in R with at least one occurrence of goal $g_i$. The score $S$ for each object is then calculated using the following formula that uses the relevance weight $rw(g_i)$ of the goal $g_i$ instead of $iof(g_i)$.

$$S_{RF}(o_j,q) = \sum_{g_i \in q} rw(g_i) gf(i,j)$$

Probabilistic RF compares the frequency of occurrence of a goal in the objects marked as relevant with its frequency of occurrence in the whole object in the SN. If a goal occurs more frequently in the objects marked as relevant than in the whole object in the SN it is assigned a higher $rw(g_i)$ weight. Then, there are two ways of choosing the goals to add to the query: (1) adding goals whose weight is over a predefined threshold, or (2) adding a fix number of goals, for example the k goals with the highest $rw(g_i)$ weight. In the experiments reported in this paper we used the second technique. After a few tests, the number of goals to be added to the original query was set to 10.

## 5. User's Query Processing and Goal Recognition Errors

This section reports some of the results of the experimental analysis of the effects of Goal Recognition Error (GRE) in query processing.

*5.1. Effects of Goal Recognition Errors on Novice User's Query Processing*

The first experimental analysis was directed towards studying the effects of different GRE in query on the effectiveness on a KE system. The parameters configuration most commonly used in objects KE employs the *gf-iof* weighting scheme on goals extracted from objects and queries. Extracted goals are first compared with a stoplist. Goals appearing in the stoplist are removed, and the remaining goals are subject to a stemming and conflation process, in order to further reduce the dimentionality of the goal space.

We study the effects of different Goal Error Recognitions in queries on the effectiveness of the KE system using the above standard configuration. Naturally, it can be noted that the best results are obtained for the perfect transcript 0, and there is degradation in effectiveness that can be attributed to Goal Error Recognition. Higher Goal Error Recognitions cause lower effectiveness. The reader can notice that the reference effectiveness (the one obtained with the perfect transcript) is quite low. The reasons for this behaviour are due to the fact that in our experiments the KE system's parameters have not been fine-tuned for the SN used and no precision enhancement technique, like for example the use of noun phrases, is employed, as it is done in almost all systems taking part[11]. The use of these techniques, therefore, would not allow an easy analysis of the causes of the loss of effectiveness.

We also show that for Goal Error Recognitions ranging from 14% to 21% there is not much difference in effectiveness. Moreover, our KE system seems to perform better with some higher levels of Goal Error Recognition than with lower ones: this is not statistically significant. Loss of effectiveness can only be observed at over 35% Goal Error Recognition, and significant low levels of effectiveness can be found for 65% Goal Error Recognition, where the number of errors in the query is so large that what is left of the original query is not enough for the KE system to work on. We can then conclude that standard KE is quite robust to Goal Error Recognitions in queries.

In order to study further the effects of Goal Recognition Error on the effectiveness of query processing, a large number of experiments using the KE system were carried out. In these experiments some of the parameters of the KE process were changed to study their effects on the effectiveness on the query processing task in relation to the different levels of Goal Error Recognition. The effect on KE effectiveness of the removal of the stemming phase of the indexing is study. Stemming has been proved to generally improve performance in KE [17]. It seems that stemming have the opposite effect in query processing, so much that the removal of such a phase actually improves effectiveness. There is no clear explanation for this phenomenon. The effect (either positive or negative) of stemming on the query goals should be very little and should not affect the performance of a KE system, but this is not what these results show.

Another interesting phenomenon was observed when the *gf-iof* weighting scheme was substituted by a weighting scheme that only uses the *iof* weight. It was surprising to observe that the *iof* weighting scheme produced the same level of effectiveness than *gf-iof*. This is in contrast to what generally happens in KE, where the *gf* weight is very important [17, 19, 20]. Other experiments involving the use of different versions of the *gf* weighting scheme and of different sizes of stoplists did not produce significantly different results from the ones reported here.

*5.2. Effects of Goal Recognition Errors and Novice User's Query length*

To test the robustness of query processing in relation to query length, another series of experiments was conducted. In fact, it is intuitive to think that the same Goal Error Recognition would have a much detrimental effects on short queries than on long ones. We report average and median precision values for queries at different levels of Goal Error Recognition. In this study, short queries are queries with less than 16 terms, and long query those with more than 16 terms; where 16 terms is the median length of queries. The average number of terms in a query, after stopterm removal is 20, therefore there is a number of considerably long queries raising the average.

We can notice that short queries have a lower average precision for any level of Goal Error Recognition, while long queries have higher average precisions for any level of Goal Error Recognition. This proves the intuition that long queries are more robust to Goal Error Recognition than short queries. However, we should also notice that the median values for all levels of Goal Error Recognition are always better than the average values, suggesting that some queries give very bad performance and lower the average. The strange behaviour of the KE system for the 50% Goal Error Recognition, that give better performance than some lower Goal Error

Recognitions, can be explained by correct recognition of one or more important terms than enabled that run to find one or more relevant objects than other runs at lower levels of Goal Error Recognition.

## 6. Effectiveness of Relevance Feedback in Novice User's Query Processing

Given the acceptable level of effectiveness of a KE system performing User's Query Processing at levels of Goal Error Rate roughly below 30%, we can conclude that it will be quite likely that in the first n retrieved objects in the Semantic Network there will be some relevant ones. It is therefore possible to use both standard and pseudo RF to try to improve the effectiveness of the query recognition task.

### *6.1 Standard Relevance Feedbacks*

Standard RF was performed using the first n known relevant objects in the SN found in the ranked list of retrieved objects. This process simulates a novice user manually selecting the relevant objects in the Semantic Network.
The effects of using n = 1 and n = 2 relevant objects in the standard RF process is shown. Comparing results we can observe a significant increase in the effectiveness of the KE system (as measured by average precision) for every level of Goal Error Recognition. We can also notice that the difference between effectiveness of short and long queries has almost disappeared; this is particularly true when 2 objects are used in the RF. Both effects are due to the expansion of the original queries with goals extracted from the relevant object(s).

### *6.2 Pseudo Relevance Feedbacks*

Pseudo relevance feedback was performed using the first n objects in the ranked list of retrieved objects, irrespective of the actual relevance or not of the objects. This process can therefore be performed without any direct user involvement.

We study the effects of using n = 1 and n = 2 relevant objects. The same conclusions regarding the increase in effectiveness for any level of Goal Error Recognition that were reached for standard RF are also partially valid here. However, one can notice that the effectiveness has not improved as much, and that the difference between short and long queries has effectively disappeared. These effects can be explained by considering that pseudo RF may use objects that are not actually relevant. In fact, an analysis of the data proved that about 40% of the objects used by the pseudo RF where not relevant, on average.

## 7. Limitations

The work reported in this paper studies the effectiveness of standard and pseudo relevance feedback on novice user's queries processing. In the previous sections we reported some results of an extensive analysis of the effects of Goal Recognition Error in user's queries on the effectiveness of a KE system. We believe the work presented here to be complete since it uses a larger number of query sets, a larger collection of objects, and a more classical KE system that was not tuned to the test collection used.

Nevertheless, there are at least two important limitations to this study:

1. In the experimentation the queries used were not really representative of typical novice user's queries. However, it has been long recognised that novice user's query is mainly dependent upon the application domain and the KE environment.

2. The Goal Error Recognitions of the queries used in this experimentation were typical of dictated novice user's queries, since this was the way they were generated. Dictated user's query is considerably different from spontaneous one. We should expect spontaneous novice user's queries to have higher levels of Goal Error Recognition and different kinds of errors. Unfortunately, there is no set of spontaneous novice user's queries available for query processing experimentation and its construction is not an easy task.

# 8. Conclusions and futur works

This paper reports on an experimental study on the effects of Goal Recognition Error on the effectiveness of Query Processing for KE. Despite the limitations of the experimentation presented here, the results show that the use of classical KE techniques for Novice User's Query Processing is quite robust to considerably high levels of Goal Error Recognition (up to about 50%). Moreover, both standard RF and pseudo RF enable to improve the effectiveness of Novice User's Query Processing.

As a futur work, we propose that our KE System will be tested on short and long Novice User's Queries to show that experimental results prove or not the effectiveness of the approach proposed. This approach can serve as a basis for our research to elaborate a general methodology to diagnosis the purpose Goal of the subject.

| Data set | Size of the data set |
|---|---|
| Number of objects | 25 |
| Size of collection in Mb | 27 |
| Unique goals in objects | 15 |
| Average Objects length in term of attributes | 7 |
| Average object length (unique goals) | 15 |

Table 1. Caracteristics of the Semantic Network Object collection.

| Data set | Size of the data set |
|---|---|
| Number of queries | 45 |
| Unique goals in queries | 45 |
| Average query length | 20 |
| Median query length | 16 |
| Average number of relevant objects per query | 2 |

Table 2. Caracteristics of the used queries.

| Query sets | a=14 | b=15 | c=16 | d=20 | e=21 | f=35 | g=65 |
|---|---|---|---|---|---|---|---|
| Average number substitutions % | 16.7 | 17.1 | 18.0 | 21.4 | 22.3 | 32.1 | 48.5 |
| Average number deletions % | 2.3 | 2.4 | 2.4 | 2.4 | 2.5 | 3.1 | 2.8 |
| Average number insertions % | 7.2 | 7.2 | 7.8 | 9.3 | 10.5 | 12.8 | 22.0 |
| Average number errors % | 14.5 | 14.8 | 15.9 | 19.8 | 20.6 | 35.1 | 65.3 |
| Average number sentence errors % | 35.2 | 36.0 | 36.4 | 38.1 | 41.3 | 43.6 | 56.3 |

Table 3. Caracteristics of the different query sets.